\documentclass[traditabstract]{aa}
\usepackage{graphicx}
\usepackage{natbib}
\bibpunct{(}{)}{;}{a}{}{,} 
\usepackage{lscape}

\def\cone {\ifmmode{{\rm C}{\rm \small I}(^3\!P_1\!-^3\!P_0)}
     \else{C\ts {\scriptsize I}{\small$(^3\!P_1\!-^3\!\!\!P_0)$}}\fi}
\def\ctwo {\ifmmode{{\rm C}{\rm \small I}(^3\!P_2\!-^3\!P_1)}
     \else{C\ts {\scriptsize I}{\small$(^3\!P_2\!-^3\!\!\!P_1)$}}\fi}

\def\tex {\ifmmode{{T}_{\rm ex}}\else{$T_{\rm ex}$}\fi}
\def\tmb {\ifmmode{{T}_{\rm mb}}\else{$T_{\rm mb}$}\fi}
\def\ci     {\ifmmode{{\rm C}{\rm \small I}}\else{C\ts {\scriptsize I}}\fi}
\def\hi     {\ifmmode{{\rm H}{\rm \small I}}\else{H\ts {\scriptsize I}}\fi}
\def\hh     {\ifmmode{{\rm H}_2}\else{H$_2$}\fi}

\def\ts     {\thinspace}
\def\kms    {\ifmmode{{\rm \ts km\ts s}^{-1}}\else{\ts km\ts s$^{-1}$}\fi}
\def\msol   {\ifmmode{{\rm M}_{\odot}}\else{M$_{\odot}$}\fi}
\def\lsol   {\ifmmode{{\rm L}_{\odot}}\else{L$_{\odot}$}\fi}
\def\zsol   {\ifmmode{{\rm Z}_{\odot}}\else{Z$_{\odot}$}\fi}

\usepackage{amstext}

\begin{document}

\title{\textit{MAMBO} image of the debris disk around $\epsilon$ Eridani : robustness of the azimuthal structure}

\subtitle{}

\author{
Jean-Fran\c cois Lestrade  \inst{1}
\and
Elodie Thilliez \inst{2}
}
\institute{
Observatoire de Paris - LERMA, CNRS, 61 Av. de l'Observatoire,  F-75014, Paris, France
\and
Centre for Astrophysics \& Supercomputing, Swinburne University of Technology,
Hawthorn, VIC, 3122, Australia
}

\offprints{J-F Lestrade, \email{jean-francois.lestrade@obspm.fr}}

\date{Received 27 November 2014 ; accepted 19 February 2015}
\titlerunning{MAMBO image of the debris disk around $\epsilon$~Eridani}

\abstract{The debris disk closest to Earth is the one around the star $\epsilon$~Eridani at a distance of 3.2~pc. 
It is the prime target for detailed studies
of a belt of planetesimals left from the early phase of planet formation other than the Kuiper Belt. 
The non-uniform ring-like structure around $\epsilon$~Eridani, originally discovered at $\lambda=850~\mu$m 
 with the bolometer camera \textit{SCUBA}, could be the signpost of unseen long-period planets 
interior to the disk that gravitationally 
interact with it through mean-motion resonances. However, the reliability of the structure  
at 850~$\mu$m, which has been debated, has not been verified with independent observations until now. We present a high signal-to-noise
ratio image of this structure at $\lambda=1.2$~mm made with the bolometer camera \textit{MAMBO} 
and compare this with the \textit{SCUBA} image.  We have found that three of the four emission clumps (NE, NW, SW) 
and the two deep hollows to the east and west are at the same positions in the \textit{MAMBO} and 
\textit{SCUBA} images within astrometric uncertainty. The SE clump is at odds, significantly brighter and more extended 
in the SCUBA than in the MAMBO images, but it is possible that this mismatch is an artifact. 
We conclude that this degree of positional coincidence provides tentative evidence that the observed structure is robust.
In addition, we present the radial brightness profile of our \textit{MAMBO} image and show that the width of the planetesimal belt around  $\epsilon$~Eridani is narrower than 22~AU, a more stringent upper limit than determined from 
previous observations. The corresponding  relative width is $0.1 \leq \Delta R / R \leq 0.4,$  which is lower than for the Kuiper Belt.}

\keywords{debris disks : circumstellar matter - planetary systems : formation - stars: planetary systems}

\maketitle



\section {Introduction} \label{intro}

A debris disk is a constituting part of a planetary system  according 
to the theory of formation and evolution of stars and planets.
It is the analog of our asteroid belt, or Kuiper Belt. 
Such a belt is formed by the planetesimals that could not be turned 
into planets through agglomeration processes at work  during 
the early  formation phase. Although these remnant planetesimals
are not directly observable when
they surround other stars, their 
ongoing erosion through mutual collisions may generate  enough dust 
to be detectable in thermal emission from mid-infrared to millimeter 
wavelengths or in scattered light \citep{Wyat08,Matt14}. 

The properties of such a belt $-$ the radius, width, sharpness of edges, as well as radial and azimuthal 
structures $-$ provide invaluable insights into possible perturbers, such as unseen planets just
interior to the inner edge that cannot be detected by the radial velocity technique because they have very long orbital periods. 
Stars exterior to the system are likewise perturber candidates  when the central star is part of a multiple system or 
is still embeddded in the open cluster of its birth during the first one hundred~Myr of its lifetime.   
   
The debris disk closest to Earth known today is the one around $\epsilon$~Eridani, a mature K2 type star (0.82 M$_{\odot}$ in mass and $\sim$~850~Myr in age) 
that is only at 3.2~pc, ranked the tenth star in distance from the Sun. 

A Jupiter-mass planet might orbit at about 3 AU  from $\epsilon$~Eridani according to  radial velocity and astrometric measurements 
\citep{Hatz00,Bene06,Angl12}, but a definitive interpretation of the measurements such as a $\text{seven-}$  year orbital period of 
this companion, is complicated by stellar activity cycles \citep{Cumm99,Metc13}   

The cold debris disk around $\epsilon$~Eridani was initially discovered with the \textit{Infrared Astronomical Satellite (IRAS)}
\citep{Gill86,Auma88} and was later imaged  at 850~$\mu$m and 450~$\mu$m  in using the \textit{James Clerk Maxwell telescope (JCMT)}
with the \textit{Submillimeter Common-User Bolometer Array (SCUBA)} \citep{Grea98, Grea05}. These images showed a ring similar  
to our Kuiper Belt with a noticeably azimuthal structure. The disk is observed almost face-on and offers a unique opportunity 
for detailed studies.
 
Short-wavelength observations of a debris disk predominantly
trace emission of fine grains that are dynamically scattered by radiation 
or stellar wind forces; these observations yield smooth images. In contrast, millimeter-wave observations predominantly
trace millimeter-sized
particles  whose spatial distribution is controlled by mutual collisions and gravity, not by pressure 
radiation \citep{Wyat99, Auge06, Theb14, Kral14}. 
Images at long-wavelength are usually more structured than at short-wavelengths, and the presence of clumps has been related 
to dynamical perturbations of unseen planets.

In this paper, we present a high signal-to-noise ratio image
of the structure around $\epsilon$~Eridani at 1.2mm acquired with the 117-channel Max-Planck Bolometer array {\it MAMBO-2} 
at the {\it Institut de Radioastronomie Millim\'etrique (IRAM)} 30-meter telescope on Pico Veleta in Spain. 
The broad caracteristics of the structures found in this new image at 1.2~mm corroborate the azimuthal structure previously 
identified in the $850~\mu$m image made  with \textit{SCUBA/JCMT} at a similar sensitivity. In addition, the width of the ring-like 
structure appears unresolved with the beam of the \textit{IRAM} 30-meter telescope ($10.7''$), which is smaller than 
the \textit{JCMT} beam.

In Sect.~\ref{sect:obs}, the observations conducted with \textit{MAMBO-2} are presented. In Sect.~\ref{sect:photo}, the total flux density 
of $\epsilon$~Eridani  measured at 1.2mm is used to study the Rayleigh-Jeans tail of the spectral energy distribution (SED).  
In Sect.~\ref{sect:azi}, the \textit{MAMBO}  and \textit{SCUBA} images and their azimuthal profiles are compared. 
In Sect.~\ref{sect:prof}, the radial brightness profile from our \textit{MAMBO} image is shown to be unresolved 
with the $10.7''$ FWHM beam of the \textit{IRAM} 30-meter telescope. 
In Sec.t~\ref{sect:discu}, we argue that kinship between the \textit{MAMBO} and \textit{SCUBA} 
images provides tentative evidence that the emission clumps of the structure are real.

\section{MAMBO observations} \label{sect:obs}

We mapped the  field around $\epsilon$~Eridani with the 
117-channel Max-Planck Bolometer array {\it MAMBO-2} \citep{Krey98} 
of the \textit{IRAM} 30-meter radiotelescope on Pico Veleta, Spain (2900~m). 
\textit{MAMBO-2} has a half-power spectral bandwidth from 210 to
290 GHz, with an effective frequency  centered on 250~GHz (1.20~mm) for thermal emission spectra.
 The 117-channel array is $240'' \times 240''$ in size, and
the bolometers are arranged in a hexagonal pattern 
with a beam separation of $22''$. This is twice the effective FWHM beam ($10.7''$) and provides an  undersampled image.

We used the standard on-the-fly mapping technique, where one map consists
of 39 azimuthal subscans of 74 sec in duration each with a scanning velocity of $\rm 5''/sec$ over $370''$ in azimuth, and  
with an incremental step of $8''$ over $312''$  in elevation, while wobbling the secondary mirror at 2 Hz by $60''$ 
or $46''$ in azimuth (the observations are split between the two throws). 
 This scanning pattern produces time streams of data that are converted 
into a fully sampled spatial map with $3.5''$ pixels. 

We conducted 14 forty-eight-minute observations of the field around $\epsilon$~Eridani 
on nine different days between November 16 and December 4, 2007, totaling 11.2 hours of on-source observations. 
Atmospheric conditions were good during these winter observations,
with typical zenith opacities between 0.1 and 0.3 at 250 GHz and low sky noise, that is,\textit{} low atmospheric fluctuations. 
The telescope pointing was checked before and after each forty-eight-minute map by using 
the same nearest bright point source (J0339-018); it
was found to be stable to greater accuracy than $3''$. 
The absolute flux calibration is based on observations of several standard calibration sources, including
planets, and on a tipping curve (sky dip) of the atmospheric opacity once every few hours. 
The resulting absolute flux calibration uncertainty is estimated to be about 10\% (rms).

\begin{figure}[h!]
\resizebox{8.5cm}{!}{\includegraphics[angle=-90] {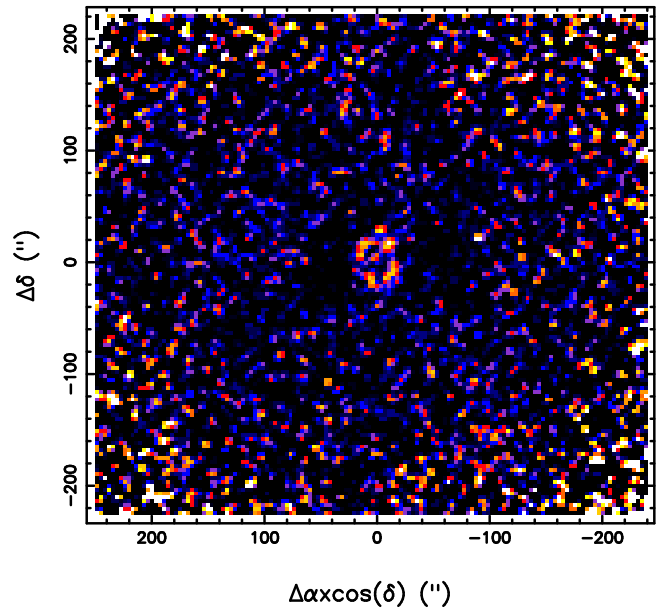}}
\caption{Intensity map of the entire field around $\epsilon$~Eridani observed at $\lambda= 1.2$~mm with \textit{MAMBO} 
in November 2007 at Pico Veleta (\textit{IRAM} 30-meter radiotelescope). 
A ring is clearly apparent in the central region of the map centered on the position of $\epsilon$~Eridani on 15 November 2007
($\alpha= \rm 3^h~32^m 55.32^s$ and  $\delta= -9^{\circ}~27'~29.7''$). The point source interior to the ring
is within one pixel from the map center coincidental with the photosphere of $\epsilon$~Eridani. 
Within $50''$ from the map center, the noise rms is 0.81~mJy/$11''$beam, and the peak brightness is 3.1~mJy/$11''$beam. 
Over the whole map, the noise increases with distance from the center because of the scanning law of the telescope. The color code
is blue, purple, red, orange, yellow, and white for increasing positive
intensity and black for all negative intensities. The pixel size is $3.5''$, and the noise is uncorrelated across pixels. 
} 
\label{fig:allfield_raw}
\end{figure}

The data were analyzed using the mopsic software package written by R. Zylka at \textit{IRAM}. The chopped observations 
acquired with the telescope pointing alternately on and off source at the wobbler rate produce  
 double-beam maps that were combined to a single map using the shift-and-add procedure 
\footnote{http://www.iram.es/IRAMES/mainWiki/CookbookMopsic} \citep{Emer79}. Compared to a complete image 
restoration, this produces maps with a sensitivity higher
by about a factor 2, at the expense of no sensitivity to emission 
structures in scan direction that are larger that the wobbler throw of $60''$ or $46''$. 
The intensity map of the entire field observed around $\epsilon$~Eridani is presented in Fig~\ref{fig:allfield_raw}.
The total integration time of $11.2$ hours yielded an rms noise of $\rm \sim 0.81~mJy/11''$~beam in the central part
of the co-added map ($r < 50''$) where the structure is located. As seen in  Fig~\ref{fig:allfield_raw}, 
noise is not uniform across the map because the  scanned field is about twice
as large as the bolometer array size, so that more data are taken in the central region of the map, where the source of
interest is, than near the edges.

\begin{figure}[h!]
\resizebox{8.5cm}{!}{\includegraphics[angle=-90] {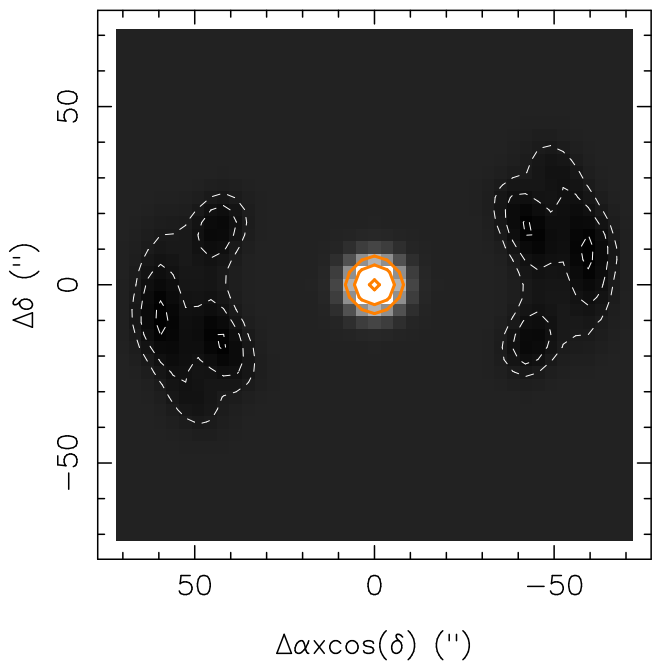}}
\caption{Synthesized beam resulting from the shift-and-add procedure used to restore the \textit{MAMBO} map from 
the chopped observations 
of the $\epsilon$~Eridani field conducted on nine days between November 16 and December 4, 2007. 
Each observation is 48 minutes long.
The hour angles of the observations comprise only $\pm 2$ hours because $\epsilon$~Eridani
culminates at the relatively low elevation of $43^{\circ}$ at Pico Veleta. Dashed white contours are -7\%, -4\%, -2\%,
orange contours are +25\%, +50\%, +90\%, of the peak at the center.
} 
\label{fig:dirty}
\end{figure}

\null
\null 

\begin{figure}[h!]
\resizebox{12.5cm}{!}{\includegraphics[angle=-90]{Fig3.ps} }
\caption{\textit{MAMBO} intensity map of the structure around  $\epsilon$~Eridani at $\lambda= 1.2$mm restored with the CLEAN algorithm.
  The star $\epsilon$~Eridani is at the map center (0,0), which is at  $\alpha= \rm 3^h~32^m 55.32^s$ and  $\delta= -9^{\circ}~27'~29.7''$ (15 November 2007).
 The telescope beam is shown as a hatched circle (FWHM = $10.7''$). 
 Contour levels are 30\%, 60\%, and 90\% of the northwest brightness peak. The first contour (30\%) is at the level 
 of the CLEAN residuals, which were not added back to the map for clarity, as is standard.
 The point source (1.2 mJy) close to the map center has been subtracted (see text of Sect. \ref{sect:photo}).}  The pixel size is $3.5''$, and the map is unsmoothed.
 \label{fig:clean}
\resizebox{6.5cm}{!}{\includegraphics[scale=0.5, angle=0]{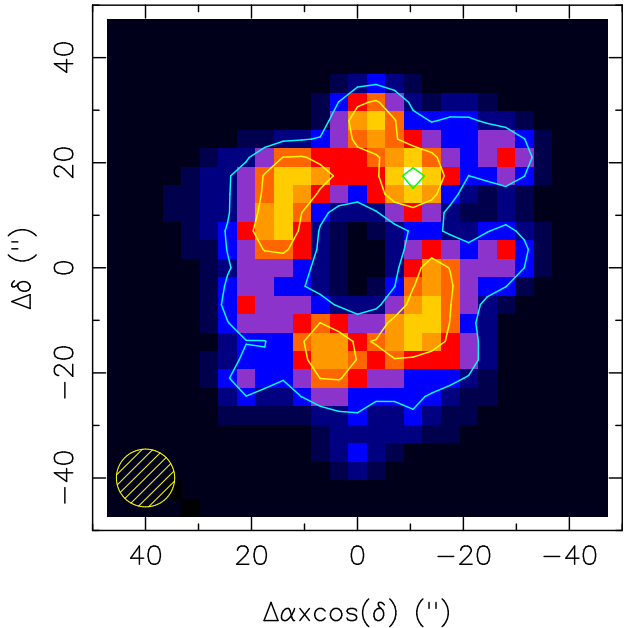} }
\caption{\textit{SCUBA} intensity image of the structure around  $\epsilon$~Eridani observed at $\lambda=850~\mu$m in \citet{Grea05}. The full field is
 $70''$ wide in RA and DEC coordinates; north is up and east to the left.} 
 \label{fig:scubaimag}
\end{figure}

In the intensity map  of Fig~\ref{fig:allfield_raw}, a ring is clearly apparent in the central region of the map, and 
a point source is located within only one pixel ($3.5''$) from the map center. 
Moreover, there are  negative (black) side lobes 
 east and west of the ring,  which have slightly distorted the outer edges 
of the structure. This is expected given that the observations were taken over 
a limited range of hour angles ($\pm 2$~hours), producing the point spread function shown in Fig.~\ref{fig:dirty} 
for the shift-and-add procedure of image restoration. To overcome these slight systematics in the map, we restored the map with 
the  CLEAN algorithm \citep{Hogb74}, which is  widely used 
for image restoration in radio interferometry with complicated point spread functions because of limited $u-v$ coverage 
\citep{Clar80}. With this algorithm, the central region of the map of Fig.~\ref{fig:allfield_raw} was
first deconvolved iteratively with the synthesized beam  of Fig.~\ref{fig:dirty}
by adopting a gain of 0.1 (fraction of the synthesized 
beam scaled to the map maximum and removed
at each iteration), and in running five hundred iterations that produced  46 CLEAN $\delta-$functions. 
Then, these resulting CLEAN spikes were convolved with a 2D Gaussian with a FWHM  
of $10.7''$ for the beam of the telescope. The final CLEAN map is shown in Fig.~\ref{fig:clean}, from which the photosphere 
has been subtracted and the residual map not added back for clarity, as is standard.

\section {Photometric excess of $\epsilon$~Eridani at long wavelengths} \label{sect:photo}

A ring is clearly apparent in the central region of  Fig~\ref{fig:allfield_raw}. Additionally, a 
point source is coincidental with  the position 
of  $\epsilon$~Eridani at the date of observation (map center) within astrometric uncertainty ($3.5''$).
The measured flux density of this point source is $1.2 \pm 0.3$~mJy. This is somewhat in excess of the flux density 
of $0.53\pm 0.25$ mJy  for the photosphere  of $\epsilon$~Eridani we predicted at $\lambda=1.2$mm with the NexGen stellar atmospheric
model \citep{Haus99}, $T_{eff}=5034~\pm~228$~K \citep{Pale15}, and the UBVRIJHK magnitudes in \citet{Duca02}.  
 
\citet{Grea14} have detected a warm component associated with $\epsilon$~Eridani in their {\it Herschel} PACS images
 that they modeled with an inner belt of 14~AU in radius ($4''$).
 The flux density of their model (their Fig.~3) extrapolated at 1.2~mm is higher than 5~mJy, meaning that {\it }
 it is well in excess of our MAMBO measurement ($1.2 \pm 0.3$~mJy),  but it may contribute to the slight excess above the
photospheric level just mentioned. Their model is based on a modified blackbody law; their emission spectrum scales 
as $\lambda^{-(2+\beta)}$  with $\beta=0.4$  longward of 150~$\mu$m. A steeper spectrum is required with our MAMBO measurement,
which indicates a population of smaller dust grains in this warm belt.

Aperture photometry made in our MAMBO image of Fig~\ref{fig:allfield_raw} has yielded a total flux density of $17.3~\pm~3.5$~mJy 
at 1.2~mm within a radius of $30''$, including the contribution
of the point source. The uncertainty includes statistical error 
and a 10\% error in absolute flux calibration. 
A first map of the structure around $\epsilon$~Eridani  at 1.2~mm was made with the \textit{SIMBA} 
bolometer array at the \textit{SEST} radiotelescope 
in Chile with a beam FWHM of $24''$. 
This map showed an extended structure that can be fit with a circular Gaussian intensity distribution with a FWHM of $36.4''$   
and a total flux density  of $21.4 \pm 5.1$~mJy \citep{Schu04},  which is consistent with our measurement at the same wavelength. 
An earlier millimeter observation  of $\epsilon$~Eridani was conducted
with an \textit{MPIfR} bolometer system also at  {\it SEST} and yielded  $24.2\pm3.4$~mJy at 1300$\mu$m \citep{Chin91}, 
somewhat in excess of our measurement, even though we observed at a shorter wavelength.          

Properties of emitting particles in the cold belt can be tested with the modified blackbody law. From the  total flux density 
of $37.0\pm2.5$~mJy  at 850~$\mu$m \citep{Grea05}  and our measurement at 1.2~mm, one finds 
$\beta=0.4^{+0.6}_{-0.4}$.  Notwithstanding the large uncertainty, this value is lower than  $\beta =1.0\pm 0.15$ 
calculated between 450 and 850~$\mu$m ($S_{450}=250 \pm 20$ mJy in \citet{Grea05}) and  might indicate a change in 
slope for the emission of medium-sized particles (pebbles).   

The compound of these slopes and uncertainties, however, is consistent with the model developed by \citet{Gasp12b} for  
a collisional cascade  that reaches 
equilibrium with a mass distribution that is steeper than the traditional solution by Dohnanyi.
Their model implies a slope of  $-2.65$ ($\beta =0.65$) 
for the long-wavelength Rayleigh-Jeans tail of the SED, consistent with our values of $\beta$ above.
The slope of their model does not depend on material properties of the medium-sized particles since they
predominantly contribute to the emission at long-wavelength in fully absorbing the incident light 
of the star ($Q_{abs}=1$), as also discussed in \citet{Wyat02}.

\section{Observed azimuthal structure around $\epsilon$~Eridani  }  \label{sect:azi}

The final \textit{MAMBO} image of the structure around $\epsilon$~Eridani at 1.2~mm is presented in Fig.~\ref{fig:clean} 
with the photosphere subtracted. The ring-like appearance is striking and was originally discovered in \textit{SCUBA}
images at 450~$\mu$m and 850~$\mu$m \citep{Grea98, Grea05}, the latter is
 reproduced in Fig.~\ref{fig:scubaimag}.

\begin{figure}[t!]
\resizebox{8.5cm}{!}{\includegraphics[angle=-90] {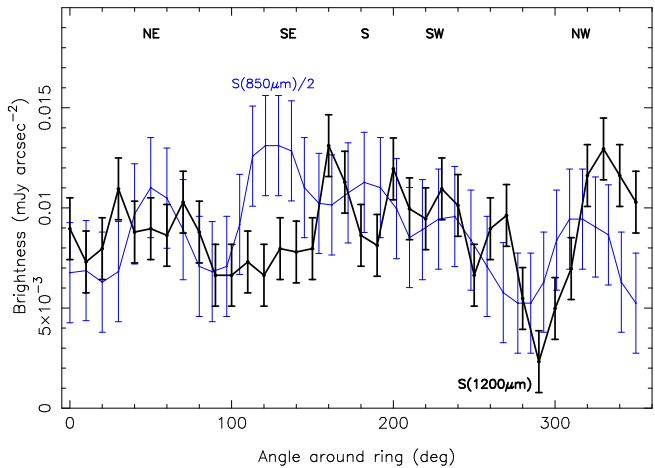}}
\caption{Azimuthal profiles of the dust emission around  $\epsilon$~Eridani computed  for \textit{MAMBO} (black) and \textit{SCUBA} (blue).
The  \textit{MAMBO} profile has been computed  from the unsmoothed image with the photosphere subtracted.  
Each \textit{MAMBO} point represents the mean brightness  over a small sector with an opening angle of 
$10^{\circ}$ and ranging radially from  $10''$ to $30''$. Points are independent 
and uncertainties are the noise rms of the map divided by the square root of the number of pixels 
in each sector ($\sim$ 6 pixels). The \textit{SCUBA} profile is 
from \citet{Grea05} and has been divided by 2, for clarity. 
The SE counterpart is inconsistent between \textit{MAMBO} and \textit{SCUBA;} this is discussed in Sect. \ref{sect:azi}.  
The angle increases counterclockwise from zero at north, 
and clumps are labeled.} 
\label{fig:azim}
\end{figure}

The main features  in the \textit{MAMBO} and  \textit{SCUBA} images  are  similar.    
There are four emission clumps  
in the southeast, northeast, northwest, and southwest sectors, and there are 
two deep hollows to the east and west (east is toward positive RA). 
These features are similarly located in the two images except for the southeast clump, 
as reported in Table \ref{tab:location}. This positional coincidence with the exception of the southeast clump 
is  also apparent by comparing  the  \textit{MAMBO} and  \textit{SCUBA} azimuthal profiles shown in Fig.~\ref{fig:azim}.

Hence, the most noticeable difference is  the southeast clump, which appears 
 as a prominent arc-like feature in the   \textit{SCUBA} image while it  is 
clearly more compact and fainter in the  \textit{MAMBO} image. 
However, in the most recent image of $\epsilon$~Eridani with \textit{SCUBA2} (Greaves and Holland, private communication),  
this clump is actually point-like and positioned
consistently with its counterpart in the  \textit{MAMBO} image.
In fact, it is possible that the arc-like feature in the  earlier \textit{SCUBA} image is an artifact that is due to
incomplete removal of atmospheric fluctuations.

The MAMBO structure we observed in November 2007 (Fig.~\ref{fig:clean}) is slightly elongated in the north-south direction, while the \textit{SCUBA} 
structure resulting from the integration of  data from 1997 to 2002 (Fig.~\ref{fig:scubaimag}) is oriented southeast to northwest.
This appears inconsistent, but the most recent image of  $\epsilon$~Eridani made in March 2011 with \textit{Herschel/PACS} 
 shows a structure that is slightly elongated in the north-south direction \citep{Grea14}, thus similar as with \textit{MAMBO} in November 2007.  
Greaves and colleagues argued that the eastwest extension seen earlier was caused by the disk moving against 
a background of contaminant sources, with the significant proper motion of the star ($1''$/yr) almost entirely westward (see also \citet{Poul06}).
Hence, these contaminants contribute to the brightness distribution of the disk, and snapshots  will eventually disentangle the 
two emissions. It is already clear from the available data, however, that the disk emission dominates.   

The slight elongation of the structure in the north-south direction is indicative of a moderately inclined structure 
($32^{\circ}$) \citep{Grea14}.  

The overall agreement between the two images made with \textit{MAMBO} and \textit{SCUBA}  provides tentative evidence 
of the presence of four emission clumps, but to characterize
them  fully, additional observations are required.

\begin{table}
\caption{Coordinates of the four emission clumps in the images.}
\label{tab:location}
\centering
\begin{tabular}{l| c c | c c}
\hline\hline
              &    \multicolumn{2}{c}{MAMBO}                  &  \multicolumn{2}{c}{SCUBA}                                                      \\ 
 Sector      &  $\Delta \alpha cos\delta$  ($''$)  &  $\Delta \delta$  ($''$)  &  $\Delta \alpha cos\delta$   ($''$)   &  $\Delta \delta$   ($''$) \\
\hline
 NW           &   $-9.5$                         &    $+18$               &    $-11$                         &   $+16$             \\
 SW           &   $-9.5$                         &    $-9$                &    $-12$                         &   $-11$            \\
 NE           &   $+16$                          &    $+9$                &    $+16$                         &   $+11$             \\
 SE           &   $+6$                           &    $-15$               &    $+18$                         &   $-13$              \\
  \hline
\end{tabular}
\end{table}

\section{Observed radial structure around $\epsilon$~Eridani }  \label{sect:prof}

\begin{figure}[h!]
\resizebox{8.5cm}{!}{\includegraphics[angle=-90] {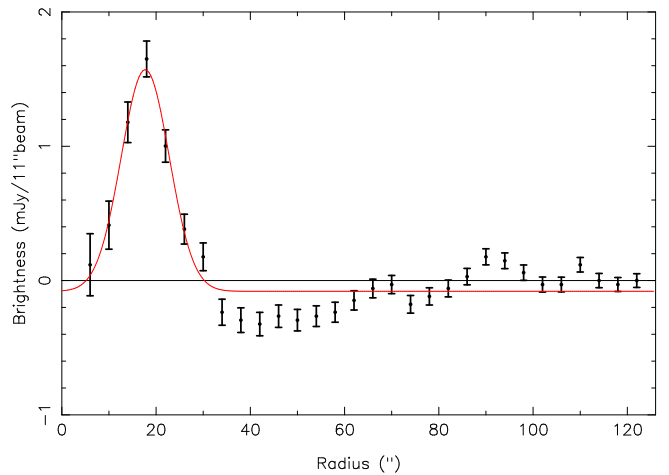}}
\caption{Radial profile of the dust emission around $\epsilon$~Eridani computed from the photosphere-subtracted
\textit{MAMBO} image.  
Points are independent and represent the mean brightness calculated over $4''$ wide, elliptical 
annuli of increasing semi-major axis (structure is inclined by 25$^{\circ}$ relative 
to the plane of the sky along the north-south direction (PA=0$^{\circ}$)). The uncertainty of each brightness point 
is the noise rms of the map divided by the square root of the number of pixels in each annulus. The Gaussian in red is based on
 the four-parameter fit described in Sect. \ref{sect:prof}.} 
\label{fig:prof}
\end{figure}

The radial brightness profile of the structure in Fig.~\ref{fig:prof}  was
calculated by averaging intensities over  $4''$ wide, elliptical annuli of increasing radii. To maximize the peak of the
profile, the structure must be projected on the sky 
with an inclination of $\sim 25^{\circ}$ and PA of $\sim 0^{\circ}$, and  
the center of the ellipses must be displaced by $2''$ westward from the map center. This optimal geometry
 is consistent with the slightly inclined structure ($32^{\circ}$) along the north-south direction seen 
 in the 160~$\mu$m image provided with {\it Herschel/PACS} \citep{Grea14}. 
 
We have fit a Gaussian  and a constant level 
($ c + a \exp\big(-0.5\times \big({(r-r_0)/\sigma_0}\big)^2\big) $) to this 
measured radial profile and found a constant level $c=-0.08\pm 0.03$~mJy/$11''$beam, an amplitude 
$a=1.65\pm 0.13$~mJy/$11''$beam, a peak radius 
$r_0=17.7\pm0.4''$, and an 
FWHM $= 2 \sigma_0 \times \sqrt{2 ln2} = 12\pm 1''$. The fit is characterized  
by a normalized $\chi^2_{\nu}$ of 0.95 for the number of degrees of freedom of 26. 
The peak radius ($17.7'' \pm 0.4''$) is comparable to  $\sim 18''$ measured in \citet{Grea05}. 

Hence, at the distance of the star of 3.2~pc, 
this peak radius corresponds to $R=57 \pm 1.3$~AU. 
The fit FWHM of the profile is consistent with the
\textit{IRAM} 30-meter radiotelescope beam  of $10.7''$ \citep{Kram13}. Thus,
we estimated limits for the belt width $\Delta R$ in deconvolving  the 
profile with the measured lower and upper limits of its FWHM,   
$11'' \leq \sqrt{10.7''^2 + (\Delta R)^2} \leq 13''$, and found $2.5'' \leq \Delta R \leq 7''$, or  $8 \leq \Delta R \leq 22$~AU at 3.2~{\rm pc}. 
Hence, the fractional belt width is $0.1 \leq \Delta R / R \leq 0.4$, 
smaller than the fractional width of the Kuiper Belt
$\Delta R / R$ of 0.5 in \citet{Ster97}, and possibly as narrow as the belt around Fomalhaut 
at millimeter wavelength (Boyles et al 2012). 

Our upper limit of the  width of the belt around $\epsilon$~Eridani 
is more stringent than the estimate made from the radial profile at 850~$\mu$m in \citet{Grea05}
because the beam of the JCMT is broader $(14'')$. Moreover, \citet{Grea14} modeled their {\it Herschel} PACS images and 
determined  a mean radius of 61~AU and a width of $\sim 15$~AU ($\Delta R / R = 0.2-0.3$),
indicating that the belt is quite radially confined according to the model. 
 
\begin{table*}
\caption{Sizes of debris disks estimated from images at long
wavelength. }
\label{tab:radii}
\centering
\begin{tabular}{l|l|c c c c c |c|c|c}
\hline\hline
            &         &  $R_{in}$      &     $R_{out}$ &    $R_{mean}$  &   $\Delta R$  &  $\Delta R/R_{mean}$ &  $\lambda$   &         &   ref.   \\
            &         &    (AU)       &     (AU)     &    (AU)       &    (AU)        &               & ($\mu$m)     &         &         \\
\hline
 Vega       & A0V     &   $< 85$      &   $> 105$    &               &                &    $>0.2$     &   1300       &   IRAM-PdB   &  1   \\   
 HD21997    & A3IV/V  &  $55\pm16$    &              &               &   $< 108$ \tablefootmark{a}    &       $< 2$  &     886      &    ALMA  & 2   \\
 Fomalhaut  & A4V     &               &              &    141.5      &  $16\pm3$      &     0.1       &   857        &   ALMA  &  3   \\
 HD107146   & G2V     &    50         &     170      &               &                &      1.1      &     880      &    ALMA &  4   \\      
 $\eta$~Corvi & F2V   &               &              &   $164\pm 4$  &   $< 60$       &      $< 0.3$  &   850        &  SCUBA2 &  5   \\
 Sun        &   G2V   &    30         &    50        &               &                &      0.5      &              &    $-$     &  6   \\ 
 $\epsilon$ Eridani & K2V &               &              &  $57\pm 1.3$  &    $< 22$      &      $<0.4$   &      1200    & IRAM-30m &  7   \\    
AU Mic      & M1V     &  $8^{+11}_{-1}$ &  $40\pm0.4$  &               &                &      1.3      &   1300       &    ALMA &  8    \\  
\hline
\end{tabular}
\tablebib{
(1)~\citet{Piet11}; (2)~\citet{Moor13}; (3)~\citet{Bole12}; (4)~\citet{Hugh11}; (5)~\citet{Duch14}; (6)~\citet{Ster97}; (7)~this work; (8)~\citet{MacG13}.
}
\tablefoot{\tablefootmark{a} angular resolution is $1.5''$, the belt is smaller than beam, hence $\Delta R <1.5''\times 72~$pc = 108~AU.
}
\end{table*}

\section{Discussion} \label{sect:discu}

\subsection{Robustness of the clumps in the MAMBO and SCUBA images}

Asymmetries and clumps observed in 
images of debris disks can be related, in theory, to dust and planetesimals trapped in resonances
with planets  just interior to the inner edge of the belt  \citep[e.g.][]{Liou96, Ozer00, Wyat03, Kriv07}. 
However, reliable identification of predicted structures in real images at long wavelength has been difficult. 
For example, \citet{Piet11} and \citet{Hugh12} could not recover the clumpy structure  around Vega
reported earlier. Similarly, \citet{Hugh12} could not confirm  the asymmetries found above $3 \sigma$ 
in the structure around HD 107146 in previous observations.

In contrast, for $\epsilon$~Eridani, the similarities  between the \textit{MAMBO} 
and  \textit{SCUBA} images discussed here provide tentative evidence that the observed structure is robust.
The four clumps may result from mean-motion resonances with an unseen planet in the system that 
is not detectable with current radial velocity measurements because
its orbital period is as long as tens or hundreds of years. 

Inferring the orbital parameters of a planet from its imprint on the structure of a disk is  an area of active research.
The ring-like structure of the solar system between 30 and 50~AU has been described as particles trapped in  
mean-motion resonances with Neptune, and the clearing within 10~AU described as particules ejected because  
of the giant planets Jupiter and Saturn \citep{Liou96,Liou99, Moro02}.

Clumps due to mean-motion resonances must be significantly enhanced to be observable, and two mechanisms have been studied. 
Dust grains can be trapped into resonances during their transport inward 
by Poynting-Robertson drag \citep{Roqu94, Kuch03, Dell05}, or resonant planetesimals and grains can be swept along 
during planet migration \citep{Wyat03, Rech06}.
First-order mean-motion resonances ($p+1:p$, $p>0$) with low $p$ are the strongest, but planet mass and orbit eccentricity 
have a significant impact on the final structure \citep{Rech06}. 

Predicting the number of observable clumps and their relative intensities requires additional constraints. \citet{Mare09} determined 
an upper limit of four Jupiter masses for any planet (1 Gyr models) inside the inner rim of the submillimeter ring 
of $\epsilon$~Eridani with  \textit{Spitzer/IRAC} data. We note also that the two hollows east and west 
seen consistently in both the \textit{SCUBA} and \textit{MAMBO} images require a planet of sub-Jupiter mass 
if arising from first-order mean-motion resonances according to several simulations \citep{Ozer00, Quil02, Rech06}.

\subsection{Narrow belt of planetesimals around $\epsilon$~Eridani}

The \textit{MAMBO} image in Fig.~\ref{fig:clean} and the radial brightness profile in Fig.~\ref{fig:prof} currently provide the most stringent upper limit on the width of the planetesimal belt around $\epsilon$~Eridani determined
directly from observations. In Table~\ref{tab:radii}, we compare this relative width ($0.1 \leq \Delta R / R \leq 0.4$) 
to the sizes of debris disks determined from millimeter or submillimeter observations that trace parent planetesimals. 
Given its mean radius of $57 \pm 1.3$~AU, the belt around $\epsilon$~Eridani has a cavity of a few tens of AU, 
which can host a large planetary system.

The inner  edge of a planetesimal belt results from the planetary formation phase when the largest planets eject 
smaller bodies into the Oort cloud \citep{Gold14}, and from  outward migration 
\citep{Levi03}. The outer edge is set by the competition between viscous spreading in the protoplanetary disk 
and  disruptive effect of stellar flybys  in  star forming region \citep{Roso14}. Additionally, at a later stage 
of evolution, the outer edge of the disk can be truncated further and its mass depleted as a result of other destructive 
stellar flybys for a period as long as $\sim 100$~Myr, when the central star is still in the expanding open cluster 
of its birth \citep{Lest11}. 

The belt around $\epsilon$~Eridani ($57$~AU, $0.1 \leq \Delta R / R \leq  0.4$) is narrower and at a larger distance than the present-day 
Kuiper Belt ($R_{mean}=$ 40~AU, $\Delta R / R =0.5$) \citep[][inner edge $\sim$ 30~AU and 
outer edge $\sim$50~AU]{Ster97}.  Furthermore, planetesimals most probably formed within $\sim 30$ AU 
in the solar system and were subsequently pushed outward by Neptune's 2:1 mean-motion resonance during 
its final phase of migration. This means that the entire Kuiper Belt formed closer to the Sun and was transported 
outward during the final stages of planet formation \citep{Levi03}. Hence, the original 
Kuiper Belt  around the younger Sun was more compact than it is today, contrasting even more with the relatively large belt
around $\epsilon$~Eridani, which is only 850~Myr in age \citep{Difo04}.

It is useful to compare the properties of the disk around $\epsilon$~Eridani  also with other disks that are spatially resolved 
 and are listed in Table~\ref{tab:radii}.  The broadest belt is 
around the solar-type star HD107146 with $\Delta R/R \sim 1.1$ \citep[][inner radius $\sim$ 50~AU, outer radius $\sim$ 170~AU]{Hugh11},
and the narrowest belt is around the A-type star Fomalhaut~A 
with $\Delta R/R \sim 0.1$ \citep[][mean radius 135~AU, width $16 \pm 3$~AU]{Bole12}. These authors also showed that 
two shepherding planets on both sides of the belt of Fomalhaut~A can be responsible for its narrow width.
This might be of interest for the  $\epsilon$~Eridani belt. 

Finally, our characterization of  the belt around $\epsilon$~Eridani 
with a mean radius of $57\pm 1.3$~AU and a width smaller than $<22$~AU
is not consistent with model fits to the SED of this star. \citet{Back09} found a broader submillimeter ring between 35 and 90 AU,
and \citet{Reid11} found a  birth ring between 55 to 90~AU. In this latter model dust grains are transported inward by 
 Poynting-Robertson and stellar wind drags and are constrained by the  $850~\mu$m flux density from JCMT/SCUBA. 
Unfortuntately, the authors of these two studies  
did not predict the flux density of the disk at 1.2~mm for comparison with our MAMBO measurement.

\null
\null

\section{Conclusion} \label{sect:concl}

The  \textit{MAMBO} image of the debris disk around $\epsilon$~Eridani at 1.2~mm we presented shows  a structure similar 
to the one identified earlier with \textit{SCUBA} at 850~$\mu$m : a ring-like structure broken into four
 emission clumps in the northeast, northwest, southwest, southeast sectors, and two deep hollows east and west. 
In theory, this structure can be the imprint 
of resonant clumps related to gravitational perturbations from undetected long-period planets in the system. 

 However, the reliability of images made with a single radiotelescope and bolometer camera has  been debated before because these images can be distorted if the signal of the large atmospheric fluctuations 
 inherent in this type of observation is not completely removed from the data. 

We argued that the identification of three of the four emission clumps (NE, NW, SW) 
at the same locations within astrometric uncertainty in two images made 
with different radiotelescopes, namely the \textit{IRAM} 30-meter and  \textit{JCMT},
and different bolometer cameras, namely \textit{MAMBO} and  \textit{SCUBA}, provides tentative 
evidence that the observed structure is robust. The southeast clump is the most discrepant 
in brightness and extension in the two images, but it is possible that this is an artifact.
Overall, this provides new impetus for future studies 
to relate this structure to the presence of  undetected planets in the system. 

 Additionally, we provided the most stringent upper limit on the width of the belt of planetesimals 
around $\epsilon$~Eridani with $8 \leq \Delta R \leq  22$~AU. This corresponds to a relative width of $0.1 \leq \Delta R / R \leq  0.4,$ which
is narrower than for the Kuiper Belt.  

 Future observations at long wavelengths on finer scales with interferometers will provide additional information
 on the emission clumps in the belt around $\epsilon$~Eridani, for instance,\textit{} their true size,  to investigate
 the interplay between debris disks and associated planetary systems.

\null
\null

\begin{acknowledgements}
We are grateful to the staff of the IRAM 30-m telescope, especially St\'ephane Leon now at ALMA,
for his dedication in managing the MAMBO pool, to Patrick Charlot 
at Observatoire de Bordeaux for useful
discussions on the CLEAN algorithm, and to Sarah Maddison at Swinburne University for her benevolent assistance.
We are very grateful to our referee for constructive suggestions that have improved the paper. This work was funded in part by
CNES and CNRS PNP.
\end{acknowledgements}

\null
\null

\bibliographystyle{aa}
\bibliography{25422_ap}



\end{document}